\documentclass[final,5p,times,twocolumn]{elsarticle}

\journal{Nuclear Instruments and Methods in Physics Research A}

\usepackage{amssymb}
\usepackage{graphicx}
\usepackage{epstopdf}
\usepackage{siunitx}
\usepackage{booktabs}
\usepackage{IEEEtrantools}
\usepackage{amsmath}
	\numberwithin{equation}{section}
\usepackage[colorlinks=true]{hyperref}
\usepackage[all]{hypcap}
\usepackage{subfigure}
\usepackage[skip=3pt]{caption}
\usepackage{lineno}

\usepackage[figuresright]{rotating}

\begin{document}
\begin{frontmatter}




\title{Preliminary RF design of an X-band linac for the EuPRAXIA@SPARC\_LAB project}



\author[sapienza,lnf]{M. Diomede\corref{mycorrespondingauthor}}
\cortext[mycorrespondingauthor]{Corresponding author}
\ead{\string\href{mailto:marco.diomede@uniroma1.it}{marco.diomede@uniroma1.it}}
\author[lnf]{D. Alesini}
\author[lnf]{M. Bellaveglia}
\author[lnf]{B. Buonomo}
\author[lnf]{F. Cardelli}
\author[cern]{N. Catalan Lasheras}
\author[lnf]{E. Chiadroni}
\author[lnf]{G. Di Pirro}
\author[lnf]{M. Ferrario}
\author[lnf]{A. Gallo}
\author[lnf]{A. Ghigo}
\author[lnf]{A. Giribono}
\author[cern]{A. Grudiev}
\author[lnf]{L. Piersanti}
\author[lnf]{B. Spataro}
\author[lnf]{C. Vaccarezza}
\author[cern]{W. Wuensch}

\address[sapienza]{Sapienza University, Piazzale Aldo Moro 5, Roma, Italy}
\address[lnf]{INFN-LNF, Via E. Fermi 40, Frascati, Italy}
\address[cern]{CERN, CH-1211 Geneva-23, Switzerland}

\begin{abstract}

In the framework of the upgrade of the SPARC\_LAB facility at INFN-LNF, named EuPRAXIA@SPARC\_LAB, a high gradient linac is foreseen. One of the most suitable options is to realize it in X-band. A preliminary design study of both accelerating structures and power distribution system has been performed. It is based on 0.5 m long travelling wave (TW) accelerating structures operating in the $2\pi/3$ mode and fed by klystrons and pulse compressor systems. The main parameters of the structures and linac are presented with the basic RF linac layout.

\end{abstract}


\end{frontmatter}


\section{Introduction}

In the framework of research and development of novel acceleration schemes and technology, the upgrade of the SPARC\_LAB test facility \cite{ferrario2013sparc_lab} at INFN-LNF is foreseen, based on a high gradient linac. High brightness electron bunches are fundamental for the successful development of plasma-based accelerators, for instance, whereas external injection schemes are considered, i.e. particle beam driven and laser driven plasma wakefield accelerators (PWFA and LWFA, respectively). Indeed, the ultimate beam brightness and its stability and reproducibility are strongly influenced by the RF-generated electron beam. 

In this scenario the SPARC\_LAB upgrade, named as EuPRAXIA@SPARC\_LAB \cite{ferrario_eaac2017}, might be one of the possible candidates to host EuPRAXIA (European Plasma Research Accelerator with eXcellence In Applications) \cite{Walker:IPAC2017-TUOBB3}. EuPRAXIA is a design study in the framework of Horizon 2020 (INFRADEV-1-2014), funded to bring together for the first time novel acceleration schemes, based for instance on plasmas, modern lasers, the latest correction/feedback technologies and large-scale user areas. Such a research infrastructure would achieve the required quantum leap in accelerator technology towards more compact and more cost-effective accelerators, opening new horizons for applications and research.

The preliminary EuPRAXIA@SPARC\_LAB linac layout is based on an S-band Gun, three S-band TW structures and an X-band booster with a bunch compressor \cite{ferrario_eaac2017}. The booster design has been driven by the need of a high accelerating gradient required to achieve a high facility compactness, which is one of the main goals of the EuPRAXIA project. The baseline technology chosen for the EuPRAXIA@SPARC\_LAB booster is X-band. The total space allocated for the linac accelerating sections is $\approx$25 m, corresponding to an active length of $\approx$16 m taking into account the space required to accommodate beam diagnostics, magnetic elements, vacuum equipment and flanges. Two average accelerating gradient options are foreseen for the X-band linac: high gradient (HG) of 57 MV/m and very high gradient (VHG) of 80 MV/m, corresponding to double the power of the HG case. The RF linac layout is based on klystrons with SLEDs \cite{farkas} that feed several TW accelerating structures. The operating mode is the $2\pi/3$ mode at 11.9942 GHz. The preliminary RF system parameters are summarized in Table \ref{tab:RF_parameters} \cite{cpi_website}. 

\begin{table}[hbt]
\center{
\caption{RF system parameters.}
  \label{tab:RF_parameters}
\begin{tabular}{lcr}
\midrule
Frequency & 11.9942 GHz\\
Peak RF power & \SI {50}{\mega W}\\
RF pulse length $t_k$ & \SI {1.5}{\micro s}\\
Unloaded Q-factor $Q_0$ of SLED & 180000\\
\bottomrule
\end{tabular}
}
\end{table}

In this paper we illustrate the preliminary RF design of the X-band booster. The single cell parameters have been calculated by electromagnetic (e.m.) simulations. On the basis of these results, the accelerating structure length and geometry have been optimized  by numerical studies. Finally, the basic RF power distribution layout has been designed. 

\section{Single cell study}\label{sec:single_cell}

\begin{figure}[htb]
\centering
\includegraphics[scale=1]{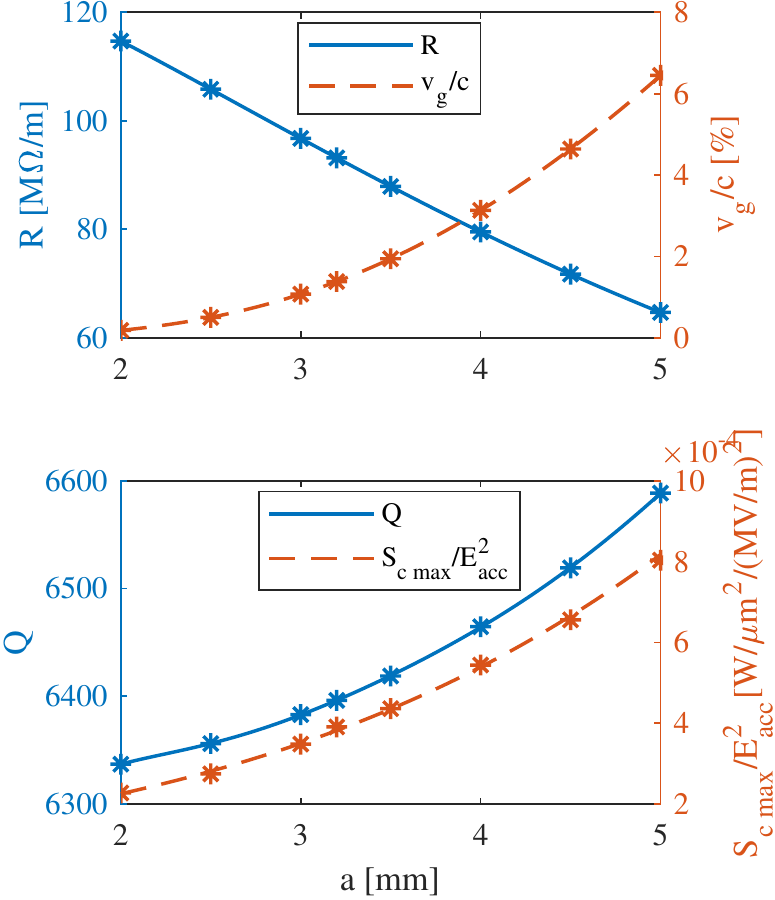}
\caption{Single cell parameters as a function of the iris radius.}
\label{fig:cell_parameters}
\end{figure}

The main single cell parameters (shunt impedance per unit length $R$, normalized group velocity $v_g/c$, Q-factor $Q$, peak value of modified Poynting vector \cite{poynting,poynting4} normalized to the average accelerating field $S_{c\;max}/E_{acc}^2$) as a function of the iris radius $a$ have been calculated with ANSYS Electronics Desktop \cite{hfss_website}. The results are reported in Fig. \ref{fig:cell_parameters}.



According to beam dynamics calculations and single bunch beam break up limits, an average iris radius $\langle a \rangle$=3.2 mm has been taken into account (the corresponding parameters are given in Tab. \ref{tab:a_3.2}) \cite{vaccarezza_eaac2017}.





\begin{table}[hbt]
\center{
\caption{Single cell parameters for an iris radius of 3.2 mm.}
  \label{tab:a_3.2}
\begin{tabular}{lcr}
\midrule
iris radius $a$ [mm] & 3.2\\
iris thickness $t$ [mm]& 2.5\\
cell radius $b$ [mm] & 10.139\\
cell length $d$ [mm]& 8.332\\
$R$ [M$\Omega$/m] & 93\\
$v_g/c$ [$\%$] & 1.382\\
$Q$ & 6396\\
$S_{c\;max}/E_{acc}^2$ [A/V] & $3.9 \cdot 10^{-4}$\\
\bottomrule
\end{tabular}
}
\end{table}

\section{Analytical optimization of structure effective shunt impedance}

The accelerating gradient distribution along the structure after one filling time $t_f$ is given by the formula \cite{anal_grudiev}:
\begin{linenomath*}
\begin{equation}\label{eq:G}
G(z,t_f) = G_0[t_f-\tau(z)] g(z),
\end{equation}
\end{linenomath*}
where $z$ is the longitudinal position, $\tau(z) = \int_{0}^{z} \frac{dz^\prime}{v_g(z^\prime)}$ is the signal time delay and $g(z)$ is defined as:
\begin{linenomath*}
\begin{equation}
g(z)=\sqrt{\frac{v_g(0)}{v_g(z)}} \sqrt{\frac{R(z)Q(0)}{R(0)Q(z)}} e^{-\frac{1}{2}\int_{0}^{z} \frac{\omega}{v_g(z^\prime)Q(z^\prime)} dz^\prime}.
\end{equation}
\end{linenomath*}
$G_0$ is the gradient at the beginning of the structure given by:
\begin{linenomath*}
\begin{equation}
G_0(t)=\sqrt{\frac{\omega R(0) P_0(t)}{v_g(0) Q(0)}}.
\end{equation}
\end{linenomath*}
$P_0(t)$ is the input RF power and, due to the SLED, is given by:
\begin{linenomath*}
\begin{equation}
P_0(t)=P_k \cdot k_{SLED}^2(t),
\end{equation}
\end{linenomath*}
where $P_k$ is the power from the klystron and $k_{SLED}(t)$ is the SLED electric field gain factor \cite{farkas}.

Integrating Eq. \eqref{eq:G} along the structure length $L_s$, we obtain the accelerating voltage $V_a$.

The efficiency of the structure is given by the effective shunt impedance per unit length $R_s$ defined as \cite{neal}:
\begin{linenomath*}
\begin{equation}\label{eq:R_s}
R_s = \frac{V_a^2}{P_k L_s} = \frac{V_a \langle G \rangle}{P_k} = \frac{\langle G \rangle^2 L_s}{P_k},
\end{equation}
\end{linenomath*}
 where $\langle G \rangle = V_a/L_s$ is the average accelerating gradient. 

\begin{figure}[htb]
\centering
\includegraphics[scale=1]{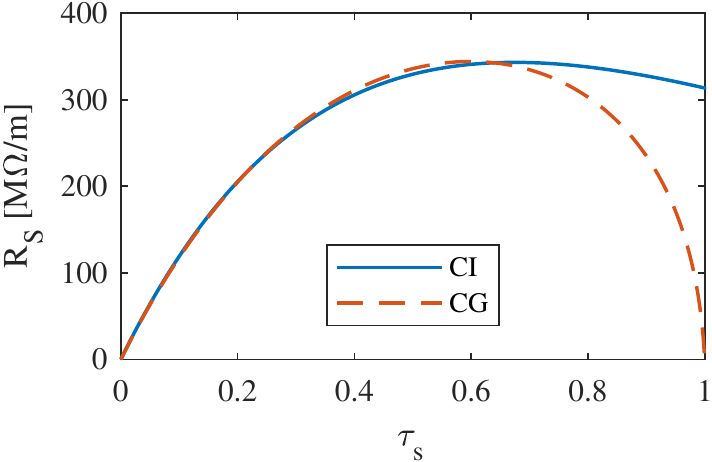}
\caption{Effective shunt impedance per unit length for the CI and CG structure.}
\label{fig:Rs_CI_CG}
\end{figure}

For a constant impedance (CI) structure $R_s$, as a function of the section attenuation $\tau_s$ ($= \frac{\omega}{2 v_g Q}L_s$), is given by \cite{leduff}:
\begin{linenomath*}
\begin{align}\label{eq:Rs_CI}
R_s =\;&2 \tau_s R \left\{ \frac{1 - \frac{2 Q_l}{Q_e}}{\tau_s} \left( 1 - e^{-\tau_s} \right) + \right.  \nonumber \\ &\left. + \frac{\frac{2 Q_l}{Q_e} \left[ 2 - e^{- \left( \frac{\omega t_k}{2 Q_l} - \tau_s \frac{Q}{Q_l} \right)} \right]}{\tau_s \left( 1 - \frac{Q_l}{Q_e} \right)} \left( e^{-\tau_s \frac{Q_l}{Q_e}} - e^{-\tau_s} \right) \right\}^2,
\end{align}
\end{linenomath*}
where $Q_l = \frac{Q_0 Q_e}{Q_0 + Q_e}$ is the loaded Q-factor of SLED (being $Q_e$ the external quality factor).
For a constant gradient (CG) structure $R_s$ is given by \cite{farkas}:
\begin{linenomath*}
\begin{align}\label{eq:Rs_CG}
R_s =\;&R \frac{2 \tau_s}{1+\tau_s} \left\{ 1 - \frac{2 Q_l}{Q_e} + \frac{2 Q_l}{Q_e} \left[ 2 - e^{-\frac{\omega t_k}{2 Q_l}} \left( \frac{1 + \tau_s}{1 - \tau_s} \right)^{\frac{Q}{2 Q_l}} \right] \cdot \right.  \nonumber \\ &\left. \cdot \frac{1 - \tau_s}{2 \tau_s} \frac{1}{1-Q/2Q_l} \left[ \left( \frac{1 + \tau_s}{1 - \tau_s} \right)^{1 - \frac{Q}{2 Q_l}} - 1 \right] \right\}^2.
\end{align}
\end{linenomath*}
Figure \ref{fig:Rs_CI_CG} shows $R_s$ as a function of $\tau_s$ for both structures, with the parameters of Tabs. \ref{tab:RF_parameters} and  \ref{tab:a_3.2}. In both cases the value of the external Q-factor $Q_e$ of the SLED has been chosen in order to maximize $R_s$. 

\begin{figure}[htb]
\centering
\includegraphics[scale=1]{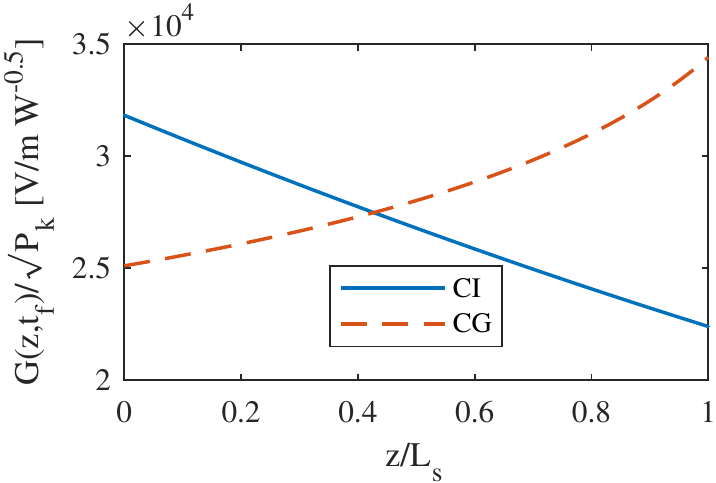}
\caption{Normalized gradient distribution along the structure for the CI and CG structure.}
\label{fig:G_CI_CG}
\end{figure}

The accelerating gradient for a CI structure is given by \cite{leduff}:
\begin{linenomath*}
\begin{align}
G(z,t_f) = &\sqrt{\frac{\omega}{v_g} \frac{R}{Q} P_k}\;e^{-\tau_s\frac{z}{L_s}} \cdot \nonumber \\ &\cdot \left\{ 1 + \frac{2 Q_l}{Q_e} \left[ e^{- \frac{\omega (L_s - z)}{2 v_g Q_l}} \left( 2 - e^{- \frac{\omega \left(t_k - t_f\right)}{2 Q_l}} \right) - 1 \right] \right\},
\end{align}
\end{linenomath*}
while for a CG structure is given by \cite{farkas}:
\begin{linenomath*}
\begin{align}
G(z,t_f) = &\sqrt{\frac{2 \tau_s}{1 + \tau_s} \frac{R}{L_s} P_k} \left\{ 1 + \frac{2 Q_l}{Q_e} \cdot \right. \nonumber 
\\ &\left. \cdot \left[ \left( \frac{1 + \tau_s \left(1 -  \frac{2z}{L_s}\right)}{1 - \tau_s} \right)^{- \frac{Q}{2 Q_l}} \left( 2 - e^{-\frac{\omega \left(t_k - t_f\right)}{2 Q_l}} \right) - 1 \right] \right\}.
\end{align}
\end{linenomath*}
The previous formulas \eqref{eq:Rs_CI},\eqref{eq:Rs_CG} allow calculating the optimum $\tau_s$ (=$\tau_{s0}$) that maximize $R_s$. This fixes also the filling time of the structure, i.e. the compressed pulse length and allows to calculate the corresponding gradient profiles given in Figure \ref{fig:G_CI_CG}.

The final optimum structures parameters are summarized in Table \ref{tab:CI_vs_CG}.

\begin{table}[hbt]
\center{
\caption{CI and CG structure parameters (analytical study).}
  \label{tab:CI_vs_CG}
\begin{tabular}{lcr}
\toprule
\textbf{Parameter} & \textbf{CI} & \textbf{CG}\\
\midrule
$R_s$ \SI {}{[M\ohm/m]} & 343 & 344\\
Optimal structure length $L_s$ [m] & 0.474 & 0.432\\
Filling time $t_f$ [ns] & 114 & 118\\
External Q-factor $Q_e$ of SLED & 20030 & 21170\\
\bottomrule
\end{tabular}
}
\end{table}








\section{Numerical optimization}

In the analytical study, the CG solution is approximated because of the assumption of constant $R/Q$ \cite{lapostolle} along the structure. On the other hand in the CI case it is quite easy to verify that, in the VHG case, one exceeds the maximum value of $S_{c\;max}$ that allows to have a Breakdown rate (BDR) lower than $10^{-6}$ bpp/m \cite{poynting,poynting4}. For these reasons we also performed a numerical study.

\begin{figure}[htb]
\centering
\includegraphics*[width=180pt]{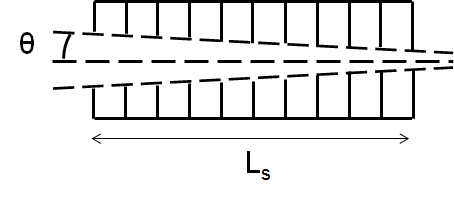}
\caption{Sketch of the linear iris tapering.}
\label{fig:tapering}
\end{figure}

To this purpose we have considered a linear tapering of the irises as sketched in Figure \ref{fig:tapering}, defined by the modulation angle $\theta$. We have then calculated (by Eq. \eqref{eq:G}) the gradient profile along the structure for different $\theta$ and $L_s$. In the calculation we have used the polynomial fits of the single cell parameters illustrated in section \ref{sec:single_cell}. From the gradient profiles we have finally calculated the effective shunt impedance per unit length and the peak value of the modified Poynting vector.

\begin{figure}[htb]
\centering
\includegraphics[scale=1]{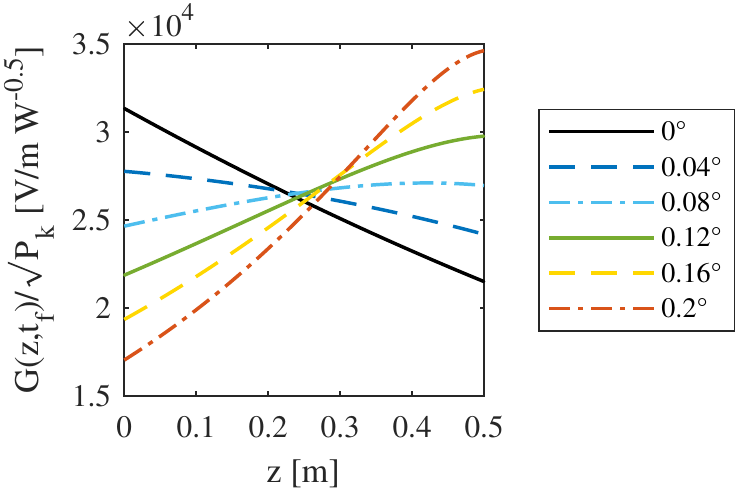}
\caption{Normalized gradient after one filling time as function of the modulation angle (0.5 m case).}
\label{fig:BFL_G}
\end{figure}

\begin{figure}[htb]
\centering
\includegraphics[scale=1]{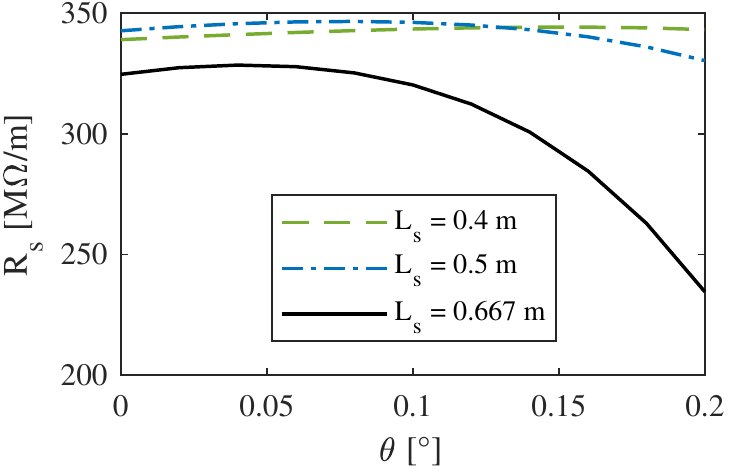}
\caption{Effective shunt impedance per unit length as a function of the modulation angle for three structure lengths.}
\label{fig:BFL_Rs}
\end{figure}

\begin{figure}[htb]
\centering
\includegraphics*[scale=1]{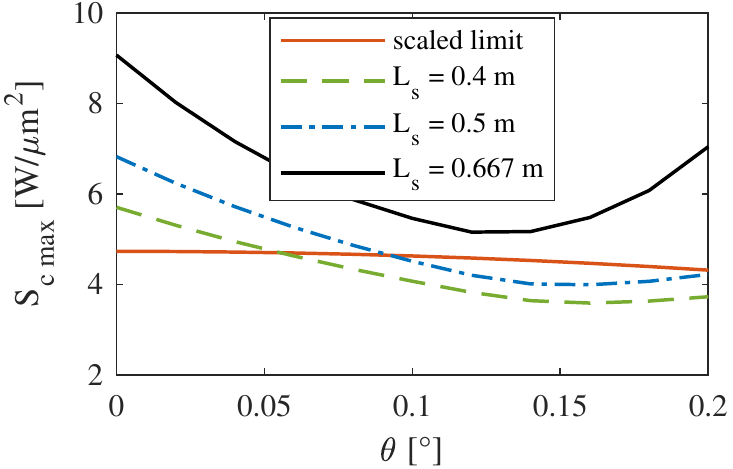}
\caption{Peak value of modified Poynting vector as a function of the modulation angle for three structure lengths (VHG case).}
\label{fig:BFL_Sc}
\end{figure}

Figure \ref{fig:BFL_G} shows, as an example, the normalized gradient profile as a function of $\theta$ for $L_s$ equal to 0.5 m. $R_s$ and $S_{c\;max}$ (VHG case), as a function of $\theta$ and for different $L_s$, are given in Figures \ref{fig:BFL_Rs} and \ref{fig:BFL_Sc}. In Figure \ref{fig:BFL_Sc} we have also reported the maximum value of $S_{c\;max}$ that, according to the scaling law given in \cite{poynting}, allows having a BDR lower than $10^{-6}$ bpp/m. From the plot it is quite easy to note that the 0.4 m and 0.5 m long structures have the same efficiency while the 0.667 m case is worse. Concerning $S_{c\;max}$, the 0.4 m solution is better but requires, on the other hand, a larger number of structures per unit length. In conclusion the 0.5 m case with $\theta$=\SI{0.1}{\degree} has been chosen as the design baseline for the X-band linac.

\section{Linac basic layout}

\begin{figure}[htb]
\centering
\includegraphics[width=180pt]{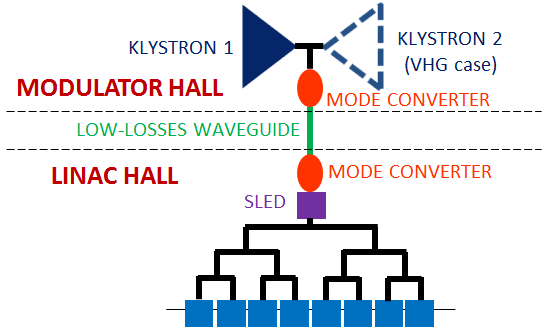}
\caption{RF power distribution layout of a single module for the HG and VHG cases.}
\label{fig:RF_layout}
\end{figure}

According to the results of numerical study, the TW X-band accelerating sections optimized for the EuPRAXIA@SPARC\_LAB application are 0.5 m long and show an effective shunt impedance per unit length of \SI {346}{\mega\ohm/m}. Commercially available X-band klystrons \cite{cpi_website} provide up to 50 MW peak power with \SI {1.5}{\micro s} long RF pulses. We have estimated the RF losses in the waveguide distribution system  of $\approx$-7 dB and, as a consequence, $\approx$40 MW available input power. The basic RF module of the EuPRAXIA@SPARC\_LAB X-band linac can be conveniently composed by a group of 8 TW sections assembled on a single girder and powered by one (for HG) or two (for VHG) klystrons by means of one pulse compressor system and a waveguide network splitting and transporting the RF power to the input couplers of the sections. The sketch of the basic module is given in Fig. \ref{fig:RF_layout} while the final main linac parameters are shown in Tab. \ref{tab:final_parameters}.

\begin{table}[hbt]
\center{
\caption{X-band linac parameters.}
  \label{tab:final_parameters}
\begin{tabular}{lcr}
\midrule
Frequency of operation [GHz] & \multicolumn{2}{c}{11.9942}\\
RF pulse length $t_f$ [ns] & \multicolumn{2}{c}{129}\\
Unloaded Q-factor $Q_0$ of SLED & \multicolumn{2}{c}{180000}\\
External Q-factor $Q_e$ of SLED & \multicolumn{2}{c}{21800}\\
$a$ first-last cell [mm] & \multicolumn{2}{c}{$3.636 - 2.764$}\\
Structure length $L_s$ [m] & \multicolumn{2}{c}{0.5}\\
Active length $L_t$ [m] & \multicolumn{2}{c}{16}\\
No. of structures $N_s$ & \multicolumn{2}{c}{32}\\
$v_g/c$ first-last cell [\%] & \multicolumn{2}{c}{$2.23 - 0.77$}\\
$R_s$ \SI {}{[\mega\ohm/m]} & \multicolumn{2}{c}{346}\\
& HG & VHG\\
Average gradient $\langle G \rangle$ [MV/m] & 57 & 80\\
Energy gain $W_{gain}$ [MeV] & 912 & 1280\\
Total Required RF power $P_{RF}$ [W] & 150 & 296\\
No. of klystrons $N_k$ & 4 & 8\\
\bottomrule
\end{tabular}
}
\end{table}

\section{Conclusions}
In the paper we have illustrated the preliminary RF design of the EuPRAXIA@SPARC\_LAB X-band linac. It has been done performing e.m. simulations of the single cell, and analytical and numerical optimization of the structure efficiency taking into account the available space, local field quantity (modified Poynting vector) minimization and available commercial klystrons. The final linac
layout and main parameters have been finally shown.

\section*{Acknowledgments}

This work was supported by the European Union's Horizon 2020 research and innovation programme under grant agreement No. 653782.




\bibliographystyle{elsarticle-num}






\end{document}